\documentclass[prl,twocolumn,showpacs,floatfix]{revtex4}

\usepackage{amsmath}
\usepackage{amssymb}

\usepackage{graphicx}
\usepackage{dcolumn}
\usepackage{bm}

\begin{document}

\title{Broken Symmetry in Density-Functional Theory: Analysis and Cure}

\author{A. Harju}
\author{E. R\"as\"anen}
\author{H. Saarikoski}
\author{M.J. Puska}
\author{R.M. Nieminen}
\affiliation{Laboratory of Physics, Helsinki University of Technology,
P.O. Box 1100, FIN-02015 HUT, Finland}
\author{K. Niemel\"a}
\affiliation{Theoretical Physics, University of Oulu, P.O. Box 3000,
FIN-90014 University of Oulu, Finland}

\date{\today}

\begin{abstract}
We present a detailed analysis of the broken-symmetry mean-field
solutions using a four-electron rectangular quantum dot as a model
system. Comparisons of the density-functional theory predictions with
the exact ones show that the symmetry breaking results from the
single-configuration wave function used in the mean-field approach.
As a general cure we present a scheme that systematically incorporates
several configurations into the density-functional theory and restores
the symmetry. This cure is easily applicable to any density-functional
approach.
\end{abstract}

\pacs{73.21.La,71.15.Mb,71.10.Pm} 

\date{\today}

\maketitle

The nanoscale semiconductor systems are technically very promising for
future components of microelectronic devices.  From the theoretical
point of view, quantum dot (QD) systems are valuable source of novel
quantum effects. Many of these result from the fact that the
electron-electron interaction and external magnetic field have greatly
enhanced effects compared to atoms and molecules.  This raises new
challenges for the theoretical methods, and the validity of
approximations in, e.g., mean-field approaches can be questioned. For
this reason, QD systems serve as perfect test cases to develop the
electronic structure methods, with the results still applicable to
great variety of physical problems where mean-field approaches have
been used.

In earlier studies, Hartree-Fock (HF) and especially
density-functional theory (DFT) methods have shown to produce accurate
results for various QD systems, even with small $N$.  However, in the
context of solutions with a broken spin symmetry, the validity of the
mean-field approaches has been actively discussed in the
literature~\cite{RM}. The spin-density wave (SDW) formation in QDs has
been compared to similar phenomena found in isotropic
metals~\cite{overhauser}, organic linear-chain
compounds~\cite{gruner}, atomic nuclei~\cite{frauendorf}, and small
fermion clusters~\cite{hakkinen}.  According to the Jahn-Teller
theorem, any non-linear molecular system in a degenerate electronic
state becomes more stable by removing the degeneracy and thus lowering
the symmetry and the total energy.  A crucial difference between
molecular and QD systems is, however, that as the nuclei in molecules
are free to move and relax, the QD potential is {\sl external} and
fixed as it results from, e.g., metallic gates.  Thus to lower the
symmetry in QD, the spin densities must ``relax'' in an
anti-ferromagnetic fashion to a SDW solution. This is claimed to
reveal the electron correlations inherent in the true ground
state~\cite{RM}.

In this Letter, we analyze symmetry breaking in a two-dimensional
rectangular QD~\cite{Esa} using both DFT and exact diagonalization
(ED).  We concentrate on the four-electron case, as it is the first
particle number showing the general features of electronic structure
seen also for larger particle numbers, such as the transitions between
the two spin states $S=0$ and 1 and the SDW solution predicted by DFT.
We find that SDW clearly reflects the limitations of basic DFT to
describe systems that have more than one major configuration in the
ground-state wave function.  There is a continuous interest for
developing DFT methods for this kind of systems. The main difficulty
for DFT is the fact that these systems have ensemble-$v$-representable
densities (E-VR) in contrast to the more common pure-state
$v$-representable densities (P-VR)~\cite{UK}.  As an interesting
feature we see a continuous transition from an E-VR to a P-VR density
as we deform our QD. Finally, we present a simple modification of DFT
that is able to describe the multi-configurational nature of the
ground states.

\begin{figure}[ht]
\includegraphics[width=\columnwidth]{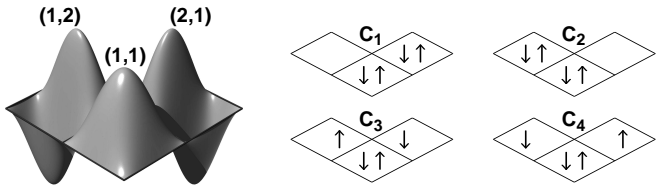}
\caption{
Left panel: The three lowest single-particle states and their quantum
numbers ($n_x$,$n_y$). Right: Electron occupations for the
four important $S_z=0$ configurations $C_i$.
}
\label{confs}
\end{figure}

The generally used model Hamiltonian of an $N$-electron QD system can
be written as
\begin{equation}
H=\sum^N_{i=1}\left[-\frac{\hbar^2 }{2m^*}\nabla^2_i+V_{\rm ext}({\mathbf
r}_i)\right] +\sum^N_{i<j}\frac{e^2}{\epsilon|{\mathbf r}_i-{\mathbf
r}_j|} \ ,
\label{ham}
\end{equation}
where we have used the effective-mass approximation to describe
electrons moving in the $xy$ plane, surrounded by background material
of GaAs with the effective electron mass $m^*=0.067m_e$ and dielectric
constant $\epsilon=12.4$. We use scaled atomic units, and energies are
thus given in ${\rm Ha^*}\approx{11.86}$ meV and lengths in
$a^*_B\approx{9.79}$ nm.  The external confinement in the $xy$ plane
is described by an infinite hard-wall potential,
\begin{equation}
V_{\rm ext}(x,y)=\left\{ \begin{array}{ll}
0, & 0\leq{x}\leq\beta{L},\,0\leq{y}\leq{L}\\
\infty, & \text{ elsewhere}.
\end{array} \right.
\end{equation}
The deformation parameter $\beta$ defines the ratio between the side
lengths of the rectangle.  The area of the dot is fixed to be $\pi^2$.
The single-particle eigenstates are sine-functions in both directions,
labeled with two quantum numbers $(n_x,n_y)$, and energies
$E_{n_x,n_y}=\left({n_x^2}/{\beta}+\beta n_y^2\right)/2$.
Fig.~\ref{confs} shows the three lowest eigenstates and the most
important $S_z=0$ configurations of the four-electron QD.

We solve the electronic structure of QD using ED and DFT. In ED, the
many-particle wave function is constructed as an expansion of the
non-interacting eigenstates.  The results approach the exact ones as
more terms are added to the expansion.  We use a basis of up to around
fifteen thousand configurations.  The interaction matrix elements are
calculated numerically using Gaussian integration.  In ED, all
many-body quantum effects are taken into account in an exact
fashion. In DFT, these are incorporated in a mean-field fashion as an
effective potential.  In the DFT method used, we allow different spin
densities for up and down electrons. This is necessary for $S \ne 0$,
and needed also for $S=0$ in order to find broken-symmetry solutions.
More details of the DFT method and the numerical implementation can be
found from Ref.~\cite{Esa} and references therein.

\begin{figure}[th]
\includegraphics[width=\columnwidth]{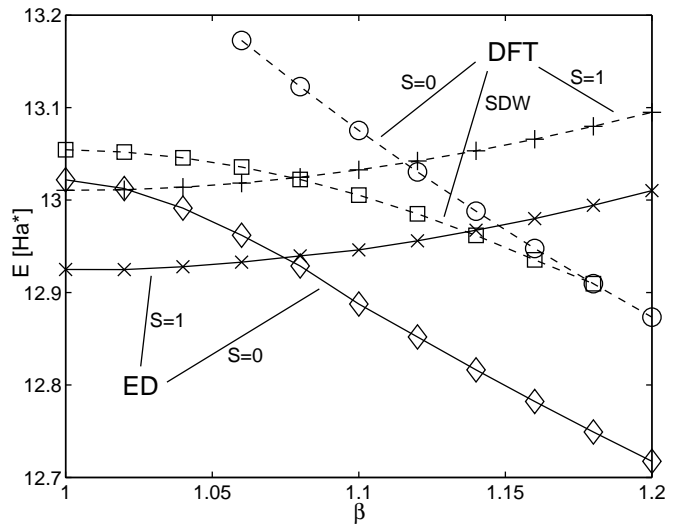}
\caption{Energy of the four-electron dot as a function of the axis
ratio $\beta$. The solid lines present ED energies, we use crosses for
$S=1$ and diamonds for $S=0$, correspondingly. The dashed lines are
DFT energies, pluses for $S=1$, boxes for the $S=0$ SDW solution, and
circles for the symmetric $S=0$ energy.}
\label{Energy}
\end{figure}

\begin{figure}[hb]
\includegraphics[width=\columnwidth]{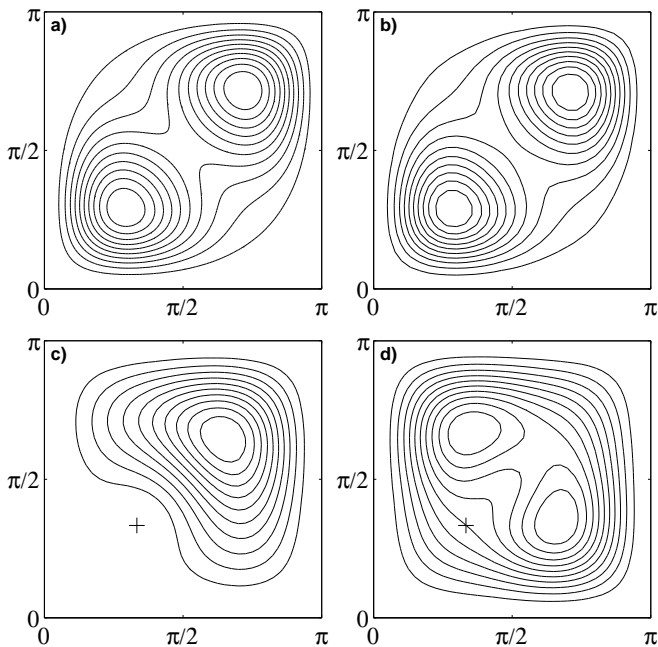}
\caption{(a): DFT spin density for the SDW solution. The density for
the other spin type can be found through rotation by 90 degrees. (b):
ED spin density for the artificial sum of $S=0$ and $S=1$ states. (c):
Conditional density from ED for the same spin type as the electron
fixed at '$+$'. (d): ED conditional density for opposite spins.  The
number of contour lines (drawn at uniform spacing) is fixed to 10 in
each figure to ease comparisons.  The SDW density is more similar to
the unphysical ED density of (b) than the conditional densities.  }
\label{dens} 
\end{figure}

In Fig.~\ref{Energy} we present the 
DFT and ED energies of the rectangular quantum dot as a function of 
the deformation parameter $\beta$.
For $\beta$ close to unity, the $S=1$ state is lower in energy than
the $S=0$ state, in accordance with Hund's rule.  In the case of
the $S=1$ state, the DFT energies compare quite well with those
obtained by ED: the deviation between the two remains nearly constant
for all values of $\beta$. Such a behavior is not seen in the $S=0$
results, for which we show two DFT energies: one with retained
symmetry and another with a broken one.  The broken-symmetry solution
has a non-zero total spin density, corresponding to a SDW solution,
see Fig~\ref{dens}.
The DFT calculation with the spin symmetry does not converge for
smaller $\beta$ than those shown. This is due to the degeneracy in the
system.  Convergence can be achieved by use of fractional occupations.
Comparing the DFT $S=0$ energies to the exact ones, one can see that,
unlike for the $S=1$ case, the error in DFT is not constant. The
energy of the symmetry-restricted state grows linearly towards $\beta
\to 1$ where the ED energy saturates. On the other hand, the SDW state
has an energy that overcompensates the error in the
symmetry-restricted energy. The energy of the SDW state is closer to
the exact value than the energy of the proper-symmetry state.  One
should note that the errors in DFT energies nearly cancel at the
ground-state transition point, and the DFT prediction for it is very
accurate.

It is claimed that the SDW spin densities reflect the internal
structure of the system~\cite{RM}.  To analyze this claim, we have plotted the
SDW spin density of the DFT and ED conditional densities in
Fig.~\ref{dens}. The conditional density is defined to be the electron
density of the remaining three electrons as the coordinates of one of
the electrons are fixed.  In addition, we plot the ED spin density for
the sum of the $S=0$ and $S=1$ states.
One can see from the densities that there is a clear
anti-ferromagnetic order in the system. Densities for parallel spins
are localized in the opposite corners. Apart from this fact, the
similarity of the conditional densities to the SDW density of
Fig.~\ref{dens} is marginal.  However, the similarity of the SDW
density to the unphysical mixture of the two spin states is very
clear.  The only difference is that the DFT density is slightly more
localized.  One should note that this similarity of the SDW solution
to a mixture of two different spin states is pointed out by Hirose and
Wingreen using ED in restricted basis~\cite{HW}.

To understand the electronic structure of the system and to analyze
the problem associated with the SDW solution, it is enough to consider
only the most important configurations in the ED solution, presented
in Fig~\ref{confs}.  The $S=1$ state, the ground state for small
$\beta$, has three different $S_z$ states which are degenerate in
energy. The one with $S_z=0$ consists of configurations $C_3$ and
$C_4$ with equal weights for all values of $\beta$. The $S=0$ state is
the ground state for large $\beta$, and it consists of the
configurations $C_1$ and $C_2$. For $\beta=1$, these have equal
weights, but for larger $\beta$, $C_2$ moves higher in energy and has
a smaller weight in the exact wave function. For $\beta\approx 1.2$,
$C_1$ is clearly the dominating configuration. One should note that at
this value of $\beta$, proper symmetry is restored in the DFT
solution. The most natural reason for the occurrence of the SDW
solution is that the basic DFT is unable to take into account more
than one important configuration for the construction of the Kohn-Sham
orbitals and the resulting densities.  In terms of the configurations,
the DFT spin densities at $\beta=1.2$ correspond to $C_1$. For smaller
$\beta$, however, the SDW spin densities can only be obtained by a
linear combination of all four configurations.  For $\beta=1$, this
linear combination is equal to the unphysical mixture of two different
spin states used for Fig.~\ref{dens} (b) above.

\begin{figure}[ht]
\includegraphics[width=\columnwidth]{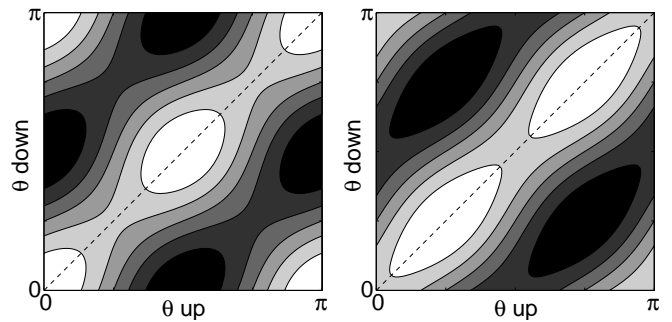}
\caption{ Single-configuration energy as a function of the two angles
in the wave function. The left panel corresponds to $\delta=0.2$, and
the right one $\delta=0.8$.  Black areas are the lowest in energy. The
proper symmetry of the wave function is found on the dashed diagonal
line.  The broken-symmetry energy minima of $\delta=0.2$ correspond to
a single configuration, and for $\delta=0.8$ to SDW solutions.}
\label{MF} 
\end{figure}

It is possible to analyze the broken-symmetry solution more generally
by considering a mean-field-type single-configuration wave function
for two up and two down-spin electrons, occupying the orbitals
$\psi_0$ and $\sin(\theta_\sigma) \psi_1 + \cos(\theta_\sigma)
\psi_2$, where $\theta_\sigma$ contains the variational freedom for a
spin type $\sigma$.  Expanding this wave function results in four
configurations similar to $\{C_i\}_{i=1}^4$ above. Assuming a further
similarity to the QD case for $\beta=1$, one can write a Hamiltonian
matrix of the four configurations as
\begin{equation}
H=\left(
\begin{array}{cccc}
E_1 & \delta & 0 & 0 \\
\delta & E_1 & 0 & 0 \\
0 & 0 & E_0 & \delta \\
0 & 0 & \delta & E_0
\end{array} \right) \ ,
\end{equation}
where the configurations couple via the off-diagonal matrix element
$\delta$ (taken to be real).  The four exact energies are $E_0 \pm
\delta$ and $E_1\pm\delta$. One can set without loss of generality
$E_0=0$ and $E_1=1$.  The single-configuration energies have an
interesting dependence on $\delta$, shown in Fig.~\ref{MF}.  We
present the energy as a function of the two variational angles
$\theta$ for cases $\delta=0.2$ and 0.8.
For small $\delta$, the second orbital for the minimum-energy solution
is $\psi_1$ for one spin type and $\psi_2$ for the other.  For the
case $\delta=0.8$, the minima are found with orbitals $\psi_1+\psi_2$
and $\psi_1-\psi_2$. The resulting total wave function of this case
can easily be found to be a sum of the two exact wave functions (with
energies $-\delta$ and $1-\delta$), and the energy of the mean-field
state is equal to the average of the two exact energies.  Furthermore,
if one assumes that $\psi_1$ has a node on the $x$ axis and $\psi_2$
on the $y$ one, one can find densities similar to the SDW solution
above.  Now for the QD, the value of $\delta$ is close to 0.8, and one
can understand the occurrence of the SDW solution from this more
general argument.  

The SDW solutions of Ref.~\cite{Esa} for larger particle numbers can
equally well be understood based on the four-electron case and the
general argument presented above.  In the cases where two Kohn-Sham
orbitals are degenerate, we have a $S=1$ ground state.  When the
aspect ratio $\beta$ is changed, the energies split and one always
finds a broken-symmetry SDW solution. The similarity to the
four-electron case follows from the fact that in all these cases,
there are two spatial orbitals of both spin type occupied by two
electrons.  The spin density in the SDW structure can then be directly
found from the two nearly degenerate states as
$(\psi_1+\psi_2)^2-(\psi_1-\psi_2)^2$.  For example, the densities in
Fig.~6 of Ref.~\cite{Esa} are accurately reproduced by this formula
using for the degenerate states the non-interacting ones with quantum
numbers $(1,3)$ and $(3,2)$ for the left panel, or $(3,2)$ and $(4,1)$
for the right panel.

Based on the results presented above, it is clear that standard DFT is
not able to describe accurately E-VR systems. The method of Ullrich
and Kohn~\cite{UK} is one possible solution, but this method might
have an underlying problem. Namely, even in the case of an open shell
and degeneracy, there are systems that still are P-VR, simply because
the configurations do not necessarily mix even if they are
degenerate. One such example is the parabolic QD, where the angular
momentum is a good quantum number and single-particle states can be
chosen in such a fashion that only one major configuration is found.
It is not straightforward to see how the method of Ref.~\cite{UK}
assort the open-shell cases that are E-VR from those that are still
P-VR.

As a possible solution we propose a scheme where first a standard DFT
calculation is performed for the system (without symmetry breaking but
with fractional occupations for the degenerate levels). For our case
with $\beta=1$, the occupations of the two highest orbitals $\psi_1$
and $\psi_2$ are $1/2$ and the DFT energy is 13.26.  One can construct
two $S=S_z=0$ DFT configurations that have density equal to the DFT
one by defining new orbitals $\phi_{\pm}=\psi_1 \pm i \psi_2$.  Now
the configurations involve the core DFT orbital $\psi_0$, and either
$\phi_+$ or $\phi_-$. The occupied orbitals in configurations are the
same for both spin electrons, similarly to $C_1$ and $C_2$ in
Fig.~\ref{confs}. The coupling of these two DFT configurations can be
approximated by
 \begin{equation}
 \delta=\int 
 \phi_+^*(\mathbf{r}_1) \phi_+^*(\mathbf{r}_2) 
 \frac{1}{r_{12}}
 \phi_-(\mathbf{r}_1) \phi_-(\mathbf{r}_2) 
\  d \mathbf{r}_1 d \mathbf{r}_2
\ .
\label{delta}
 \end{equation}
Now the DFT energy gives the diagonal Hamiltonian matrix elements and
$\delta$ the off-diagonal ones. The two-ensemble DFT energy can be
found by diagonalizing the Hamiltonian matrix.  For our example, the
absolute value of delta is found to be $\approx$ 0.14. Thus the mixing
of the two configurations lowers the DFT energy to 13.12. This value
is consistent with the ED one, if one takes into account the
difference in DFT and ED energies for the non-degenerate cases.  One
should note that for a parabolic QD, the absolute value of $\delta$ is
zero (resulting from the rotational symmetry), meaning that the
configurations (with different angular momentum) do not mix. This also
shows that our scheme correctly predicts the system to be P-VR. A
similar behavior can be found for the multiplets in open-shell atoms.

The underlying idea of the scheme presented above is that DFT is able
to efficiently describe correlation effects in a certain subspace of
the full Hilbert space. This sub-space is related to a one DFT
configuration. In P-VR cases this is sufficient for the accurate
description of the system, but for an E-VR case, there are two or more
subspaces relevant for the ground state, and DFT is unable to couple
these. This coupling can be introduced, and one natural way is via
$\delta$ of Eq.~(\ref{delta}) above.

The generalization of the scheme for cases without an exact degeneracy
of the DFT orbitals is straightforward. In addition, the approximation
made for $\delta$ can be directly used for cases with larger particle
numbers, too.  This is because the states that are occupied in both
configurations do not appear in the formula for $\delta$.  We believe
that the presented approach shows to be useful for many applications
of DFT, especially for molecules, where the calculations of chemical
reactions have observed similar problems of basic DFT~\cite{H2H2}.
More details and comparisons with other ensemble DFT approaches are
left for forthcoming studies.

Concluding, have shown that the use of a single-configuration wave
function in a mean-field theory can lead to an unphysical solution
with a broken symmetry. In our case of a four-electron rectangular QD,
the energy of the SDW solution is reasonable, but the spin densities
have only a minor similarity with the exact total or conditional ones.
We also present an analysis with a more general Hamiltonian matrix and
we feel that our findings are relevant for a great variety of systems
studied by the mean-field approaches, DFT in particular.
As a cure, we propose a scheme for incorporating systematically
several configurations into a mean-field approach. The method
presented avoids the necessity of symmetry breaking, and has a built-in
criteria to determine if several configurations are actually needed or
not.

\begin{acknowledgments}

We thank R. van Leeuwen, V. Sverdlov, and E. Thuneberg for discussions
and acknowledge support by the Academy of Finland's Centers of
Excellence Program (2000-2005).
  
\end{acknowledgments}

\end{document}